\documentclass[12pt]{iopart}
\usepackage{graphics,subfigure,epsfig,graphicx}    
\usepackage{sidecap}

\begin{document}

\title{Strange quark matter in explosive astrophysical systems}

\author{Irina Sagert$^\star$, T. Fischer$^\sharp$, M.Hempel$^\dagger$, G. Pagliara$^\dagger$, J. Schaffner-Bielich$^\dagger$,
F.-K. Thielemann$^\sharp$, M. Liebend\"orfer$^\sharp$ }
\address{ $^\star$ Institute for Theoretical Physics, Goethe University, Frankfurt am Main, Germany\\
$^\dagger$Institut f\"ur Theoretische Physik, Ruprecht-Karls-Universit\"at, Heidelberg, Germany\\
$^\sharp$Department of Physics, University of Basel, 4056 Basel, Switzerland}
\ead{$^\star$sagert@th.physik.uni-frankfurt.de}

\begin{abstract}
Explosive astrophysical systems, such as supernovae or compact star binary mergers, provide conditions where strange quark matter can appear. 
The high degree of isospin asymmetry and temperatures of several MeV in such systems may cause a transition to the quark phase already around saturation density. 
Observable signals from the appearance of quark matter can be predicted and studied in astrophysical simulations. 
As input in such simulations, an equation of state with an integrated quark matter phase transition for a large temperature, density and proton fraction range is 
required. Additionally, restrictions from heavy ion data and pulsar observation must be considered. 
In this work we present such an approach. We implement a quark matter phase transition in a hadronic equation of state widely used for 
astrophysical simulations and discuss its compatibility with heavy ion collisions and pulsar 
data. Furthermore, we review the recently studied implications of the QCD phase transition during the early post-bounce evolution of core-collapse supernovae
and introduce the effects from strong interactions to increase the maximum mass of hybrid stars. In the MIT bag model, together with the strange
quark mass and the bag constant, the strong coupling constant $\alpha_s$ provides a parameter to set the beginning and extension of the
quark phase and with this the mass and radius of hybrid stars.
\end{abstract}

\maketitle
\section{Introduction:}
The future FAIR facility at GSI, Darmstadt, will explore the equation of state (EoS) of strongly interacting matter for intermediate temperatures $T$ and high baryon 
densities $n_b$
around isospin symmetry, that is for proton fractions $Y_p \sim 0.5$.
Supernovae (SNe) and binary mergers hold environments with similar conditions for $T$ and $n_b$ but with $Y_p \leq 0.3$.
As will be discussed in the scope of this article, core-collapse SNe with matter at a low value of $Y_p$ and dynamical timescales in the range of ms, provide conditions
suitable for a phase transition to strange quark matter. 
Such a scenario was recently studied in \cite{Sagert09} applying the MIT bag approach for the EoS of quark matter and using low critical densities for 
its onset. 
Simulations with different progenitor models and two different 
bag constants led to SN explosions accompanied by a significant neutrino 
burst which can be observed by present and future neutrino detectors \cite{Dasgupta09}.
In the following, we will introduce in more detail the hybrid EoS used in the above work and analyze its influence on the dynamics of the PNS evolution. 
We will discuss the compatibility with heavy ion (HI) data and pulsar observations. Furthermore we will include first order corrections from the strong 
interaction constant $\alpha_s$
and study its influence on the maximum mass of the cold hybrid star configurations. 
\section{Initial setup for the equation of state}
\begin{figure}
\subfigure{
\includegraphics[width=6.5cm]{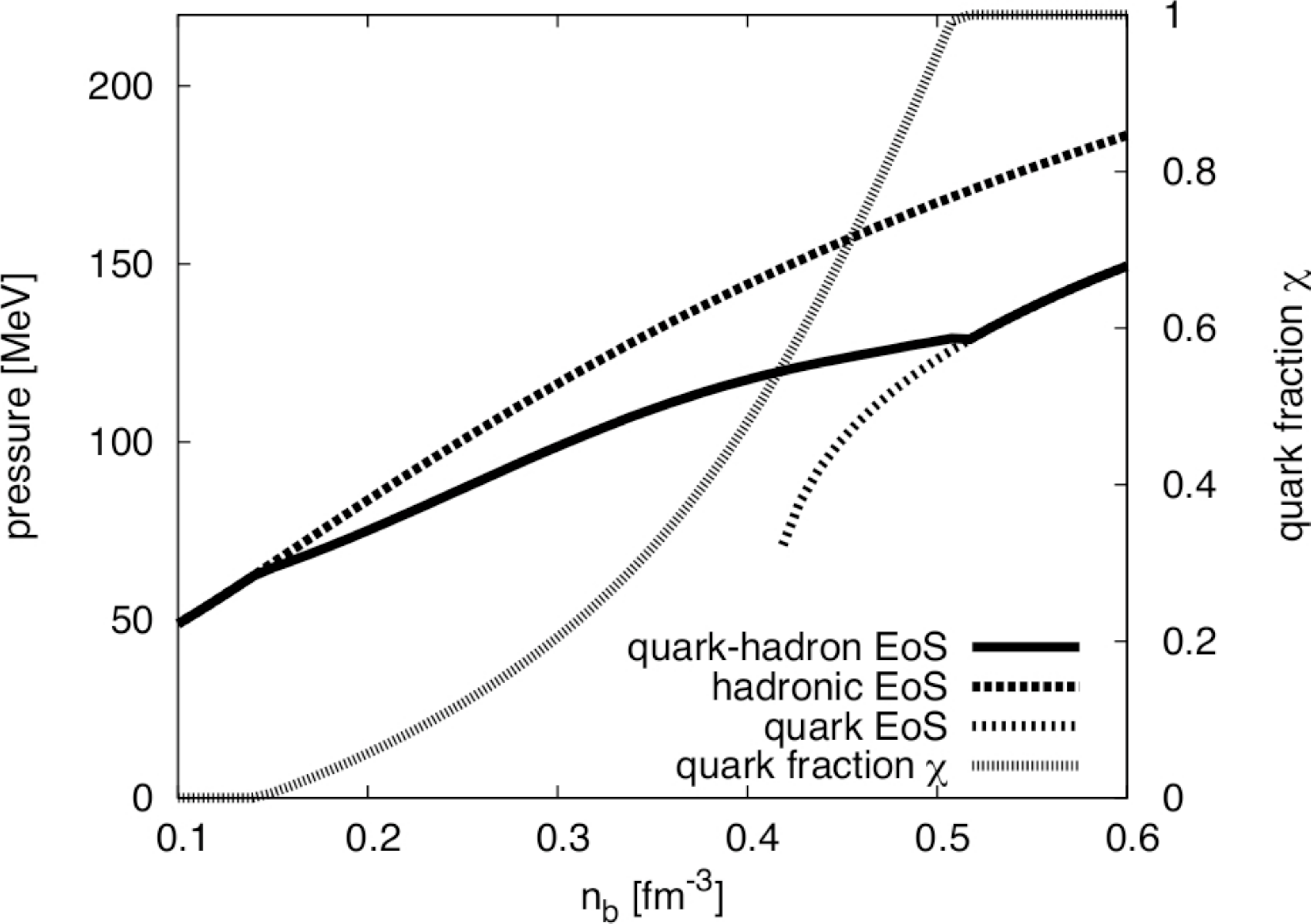}
\label{eos_yp03}}
\subfigure{
\includegraphics[width=6.5cm]{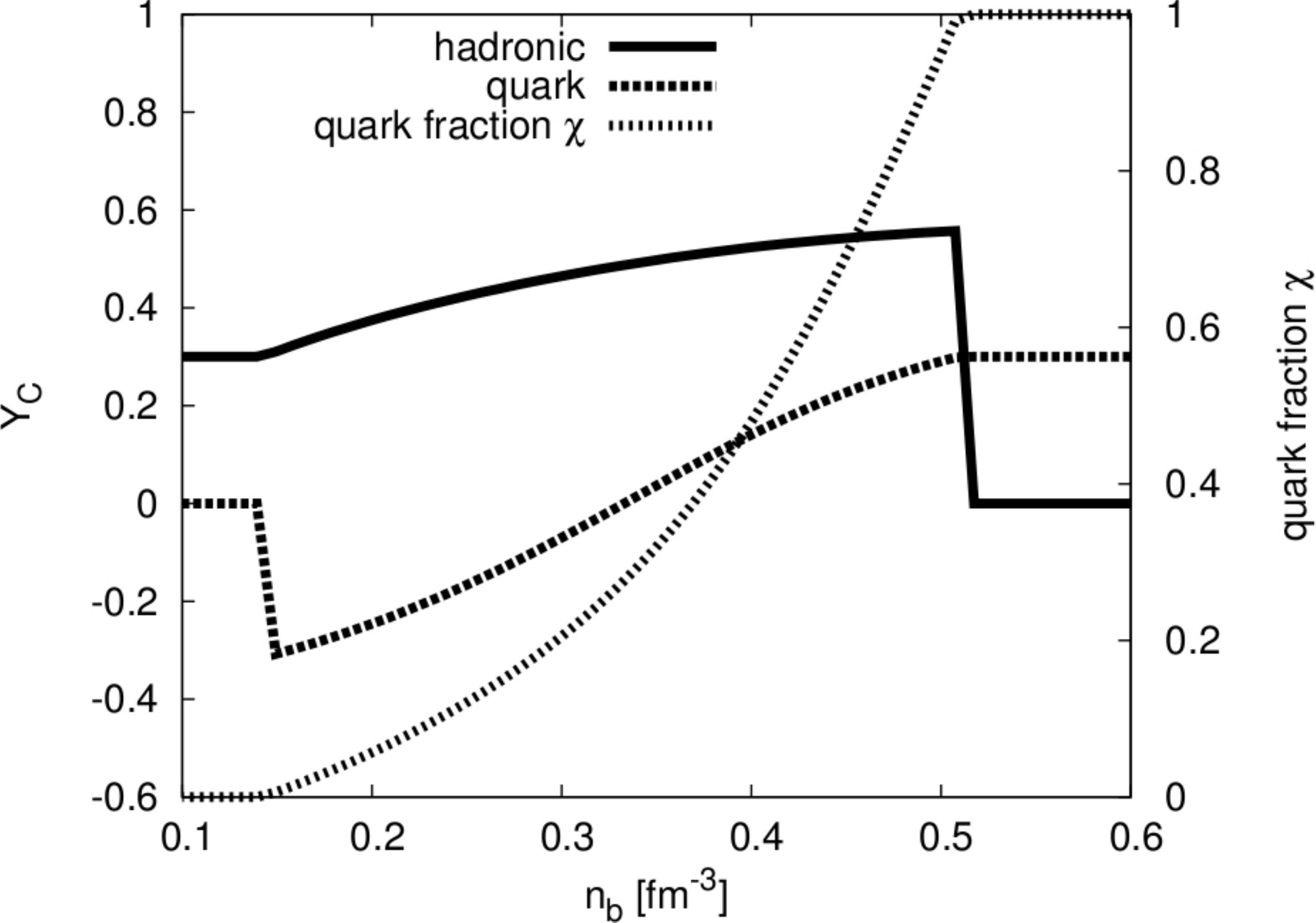}
\label{frac_yp03}}
\caption{(a) Hybrid EoS for $B^{1/4}=165$MeV and $\alpha_s=0$, together with the EoSs for the pure hadronic and quark phases at $Y_p = 0.3$ 
and $T=15$MeV;
(b) Charge fractions in quark and hadronic matter in the mixed phase for a total $Y_C = 0.3$ and temperature $T=15$MeV using $B^{1/4}=165$MeV 
and $\alpha_s=0$.}
\label{eos_total}
\end{figure}

\begin{figure}
\subfigure{
\includegraphics[width=6.5cm]{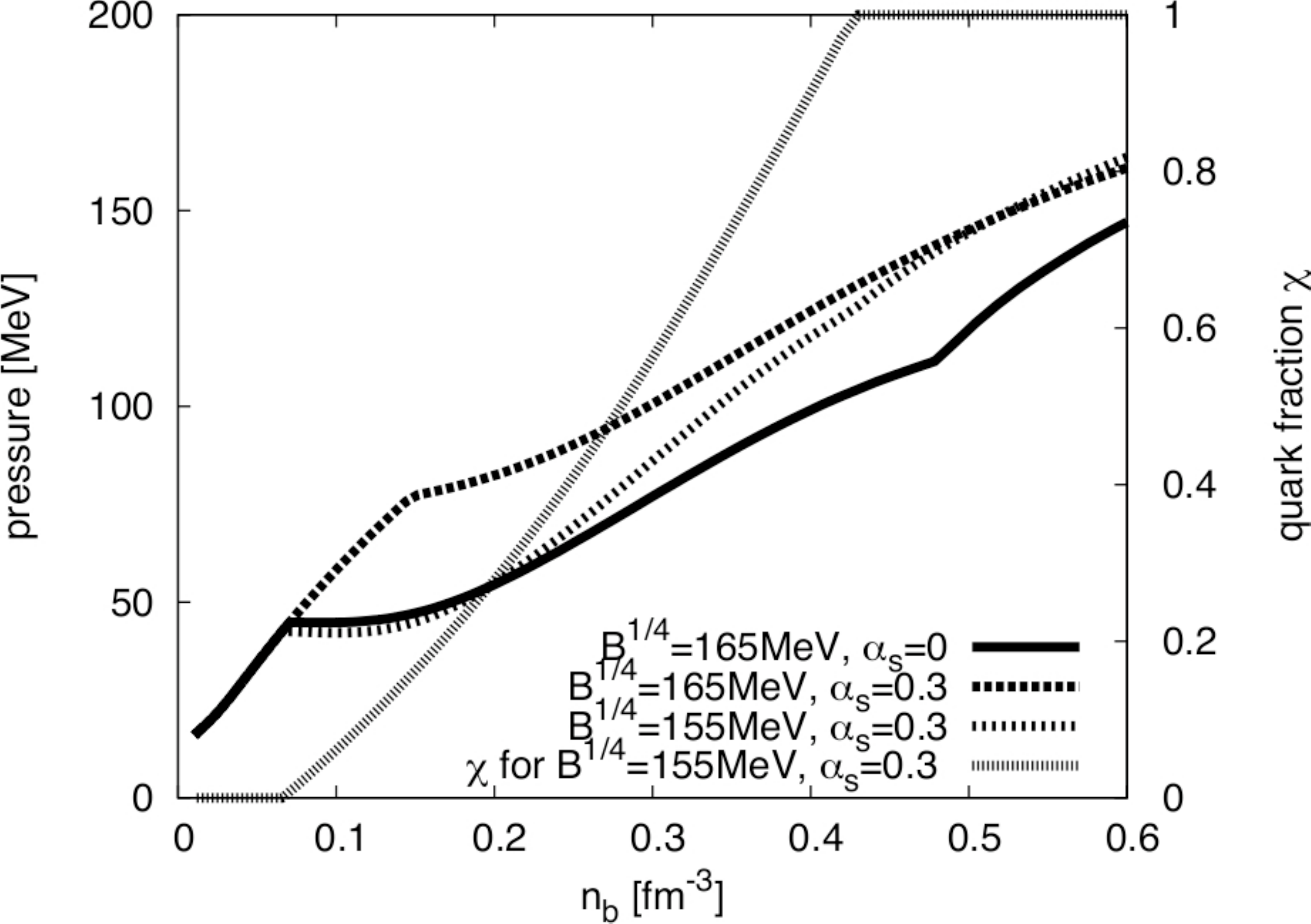}
\label{comparison}}
\subfigure{
\includegraphics[width=6.5cm]{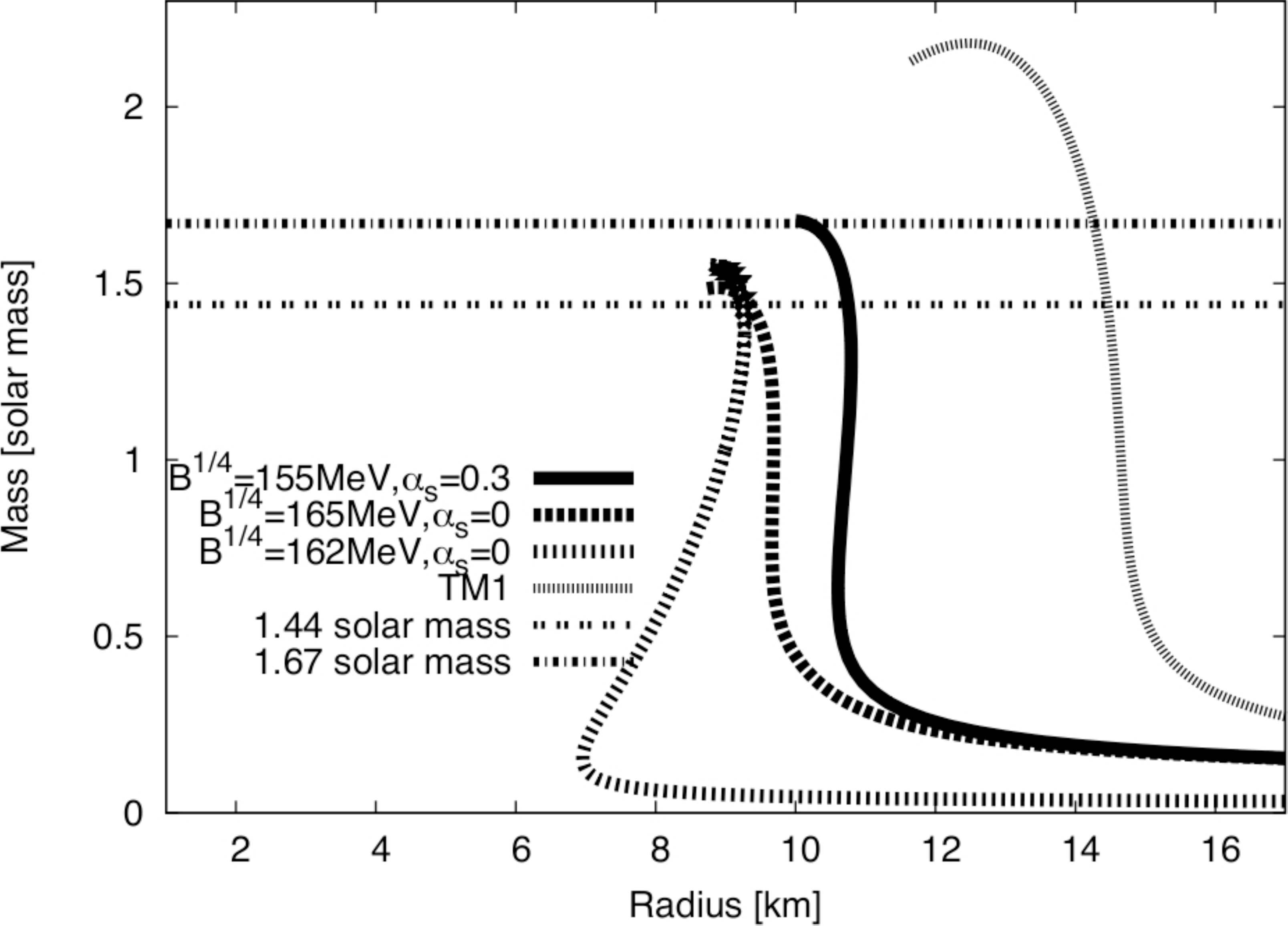}
\label{mr_as03}}
\caption{(a) The inclusion of first order corrections from the strong interaction constant $\alpha_s$ for quarks in the MIT bag model leads to an increase in 
the critical density. This can be compensated by reducing the value of $B$. The reduction in $B$ and inclusion of $\alpha_s$ results in a higher pressure in the mixed 
and quark phases
and therefore a higher hybrid star maximum mass (b).}
\end{figure}

For the hadronic part of the quark-hadron EoS we use the relativistic mean field approach by \cite{Shen:1998gq}, while quark matter is 
described by the MIT bag model. Due to their small current mass, the up and down quarks are treated as massless, while for the strange quark we 
chose $m_s=100$MeV which is well within the limits set by the Particle Data Group \cite{Amsler09}.
If quark masses are fixed and no corrections from the strong coupling constant $\alpha_s$ are included, the critical densities for the phase transition are 
directly given by the bag parameter $B$ for which we applied two values, $B^{1/4}=165$MeV and $B^{1/4}=162$MeV. 
For the construction of the phase transition we choose the Gibbs approach, where a mixed phase of quarks and 
hadrons is present \cite{Hempel09prd,Glendenning92}. 
\begin{SCfigure}
\includegraphics[width=6.5cm]{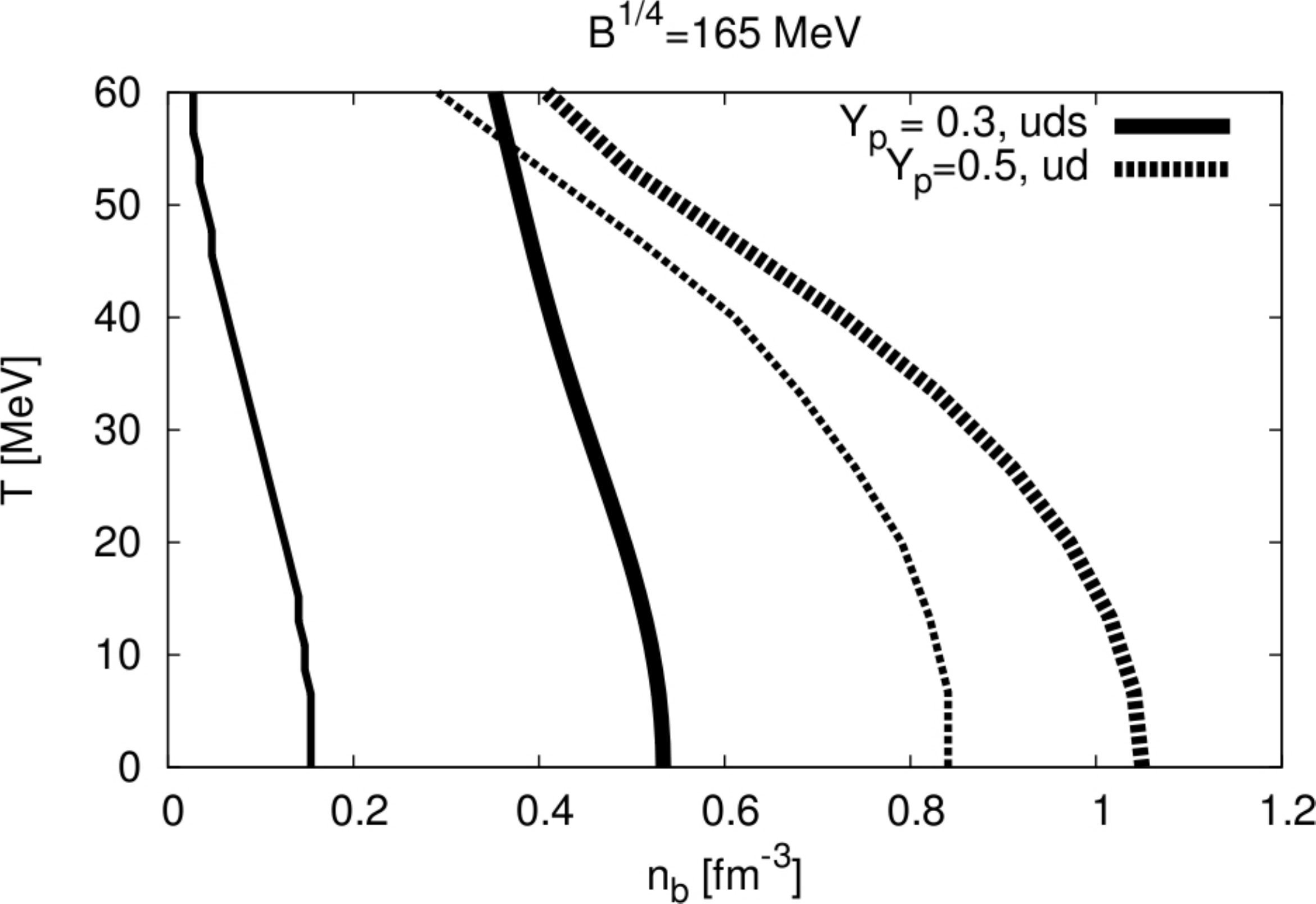}
\caption{The phase diagram for the quark matter phase transition for SN matter (solid lines) and HI collisions (dashed lines).
Thin lines denote the beginning of the mixed phase, thick lines mark the onset of the pure quark phase.}
\label{sn_hi} 
\end{SCfigure}
Figure \ref{sn_hi} shows two phase diagrams with the pure hadronic and quark phases and the mixed phase 
for $B^{1/4}=165$MeV. As can be seen, the onset of strange quark matter (denoted as \textit{uds}) for a proton fraction of $Y_p = 0.3$
happens already around saturation density $n_0$. For $B^{1/4}=162$MeV the critical densities are even smaller. 
However, these values do not contradict with results from HI collisions due to two main reasons. First, supernova dynamics happen
on timescales of ms, whereas weak processes operate within $10^{-6} - 10^{-8}$s and have therefore enough time to produce strangeness.
Consequently, phase transitions can be considered from hadronic to three flavor quark matter in weak equilibrium \cite{Mintz09}. 
For HI collisions, dynamical timescales are much shorter, of the order of $10^{-23}$s and 
therefore not long enough for strangeness to be produced and equilibrated by weak interactions. 
Consequently, for such systems, it seems to be more appropriate to consider a phase transition from hadronic to quark matter composed only of up and down quarks.
The higher the number of quark flavours and therefore the number of degrees of freedom, the softer is the EoS in the mixed phase and the lower is the critical density for its onset. 
The second main difference for SN and HI environments is the proton fraction in the two systems, being $Y_p \leq 0.3$ for the first and $Y_p \sim 0.5$ in the second 
case.
Due to the symmetry energy of hadronic matter, its isospin symmetric state is energetically favored. For a proton 
fraction $Y_p < 0.5$, the energy of hadronic matter is higher and the 
additional asymmetry pressure stiffens the EoS. This stiffness results in an earlier onset of the mixed phase with its softer EoS.
Consequently, a low value of $Y_p$ leads to smaller critical densities than for isospin symmetric matter. 
Figure \ref{sn_hi} shows two phase diagrams, for SN environments and HI collisions, for $B^{1/4}=165$MeV, illustrating that a low onset of quark matter in SN environments is compatible
with a high critical density in HI collisions, which, for the chosen $B$ and small $T$, is up to $5n_0$. 
However, the exact location of the critical density for different $T$ varies in dependence of the models for the quark and hadron EoSs, or the 
inclusion of finite size effects.\\
The softening in the mixed phase, as seen in figure \ref{eos_yp03}, is caused by the growing quark fraction $\chi$. 
Figure \ref{frac_yp03} shows the fractions of positive charge $Y_C$ in quark and hadronic matter in the pure and mixed phases. In the quark phase $Y_C$
is given by $\left(2/3 \,\, n_u - 1/3 \,\, n_d - 1/3 \,\, n_s \right)/n_b$, whereas $n_u$, $n_d$ and $n_s$ are the up, down and strange quark number densities. 
For the hadronic phase, 
the charge fraction corresponds to the proton fraction $Y_p$.
As shown in figure \ref{frac_yp03}, $Y_C$ in the quark phase can be very low and even negative.
Therefore, with increasing $\chi$, the charge fraction in the quark phase can compensate $Y_C = Y_p$ of hadronic matter, and the latter
can consequently
approach isospin symmetry towards the end of the mixed phase. At this point, due to the soft EoS in the isospin symmetric hadronic phase and the large number of 
degrees of 
freedom, the mixed phase EoS is very soft. However, the vanishing of hadronic degrees of freedom causes a significant stiffening when the pure quark phase sets in. 
Nevertheless, in the simple MIT bag model, the EoS of the pure quark phase is still much softer than the one for hadronic matter giving low maximum masses for 
hybrid stars.\\
Up to now, the highest precisely measured mass is the one 
for the Hulse-Taylor pulsar with $1.4414 \pm 0.0002$ solar masses M$_\odot$. A new candidate might be the recently studied J1903+0327, 
a millisecond pulsar with a main sequence star companion. Due to the large eccentricity of the binary system, the
advance of periastron can be measured giving a value of 1.67$\pm 0.01$M$_\odot$ for the mass of J1903+0327 \cite{Freire09}.
The mass-radius relations in figure \ref{mr_as03} show that while hybrid stars for $B^{1/4}=165$MeV and $B^{1/4}=162$MeV are
above the Hulse-Taylor constraint of 1.44M$_\odot$, their maximum masses are smaller than 1.67$\pm0.01$M$_\odot$. 
A possibility to increase the maximum masses of hybrid stars within the MIT bag model is the inclusion of first order corrections from the strong interaction 
constant $\alpha_s$ \cite{Farhi84}.
Figure \ref{comparison} shows on the example of $B^{1/4}=165$MeV how the inclusion of $\alpha_s$ corrections shifts the critical density to higher values, 
increasing the 
pressure 
in the mixed and the pure quark phases. A low density for the onset of quark matter can be obtained again by decreasing the value for $B$. 
However, the new parameter set for $B$ and $\alpha_s$ leads to a stiffer EoS in the mixed and quark phases which results in higher maximum masses of hybrid stars. 
This is 
shown in figure \ref{mr_as03} where the parameter set of $B^{1/4}=155$MeV with $\alpha_s=0.3$ leads to a hybrid star maximum mass of $\sim 1.67$M$_\odot$ and, at the same 
time, has a similar 
critical density
for the onset of quark matter as $B^{1/4}=165$MeV with $\alpha_s=0$.
\section{Quark matter in supernovae} 
We performed core-collapse simulations of low and intermediate mass
Fe-core progenitors in spherical symmetry with two different bag constants $B^{1/4}=165$MeV and
$B^{1/4}=162$MeV for the quark matter EoS. Our numerical model is based on general relativistic radiation
hydrodynamics and three flavor Boltzmann neutrino transport (for details see \cite{Liebendoerfer04} and references therein).
The conditions for the appearance of quark matter, i.e. the beginning of the mixed phase, are already obtained at the Fe-core bounce
at central densities close to and slightly above $n_0$. However, the produced small quark fraction 
initially does not influence the SN dynamics and the evolution proceeds like in a normal core collapse supernova. A hydrodynamic shock wave forms, 
travels outwards and looses energy due to the disintegration
of infalling heavy nuclei and production of neutrinos. The latter become observable in the neutrino spectra as a neutrino
burst dominated by electron neutrinos as the shock wave propagates
across the neutrinospheres (i.e. the neutrino energy and flavor
dependent spheres of last scattering). These energy losses turn the expanding shock quickly into a
standing accretion shock already $\sim 5$ms after bounce.
As discussed previously, the reason for the unchanged dynamics are the 
relative stiffness and similarity of the mixed phase EoS to the hadronic one at small quark fractions and $Y_p$ close to isospin symmetry. 
However, as matter continues to be accreted on the surface of the PNS, the density and temperature in its interior increases and a growing volume enters 
the mixed phase. 
The quark matter fraction in the mixed phase rises 
and the PNS interior moves up to softer regions of the EoS, where it becomes gravitational unstable.
A contraction proceeds into a collapse till pure quark matter is reached and the EoS stiffens again due to the disappearance of the additional 
hadronic degrees of freedom. The collapse halts and a second shock front forms at the phase boundary between the mixed and hadronic phases.
This second shock front moves outwards and turns into a shock wave when it reaches the PNS surface where the density drops over several orders of magnitude. 
Hereby, the decrease in density accelerates the shock wave to velocities of the order of the vacuum speed of light.
Shock heating of infalling hadronic matter leads to a lift of degeneracy and an increase in its proton fraction accompanied by the production of 
anti-neutrinos.
As soon as the shock wave propagates over the neutrinospheres, a second neutrino burst, dominated by anti-neutrinos is released. 
The delay of this second burst after the first deleptonization burst contains correlated information about the progenitor model, the hadronic and quark EoSs, 
and the quark-hadron phase transition. For more details, see \cite{Sagert09}.
\section{Conclusions}
Because of the different proton fractions of matter in terrestrial and astrophysical laboratories, such as the future FAIR facility at GSI and
supernovae or compact star mergers, on the one hand and their similarities in $T$ and $n_b$ on the other, 
the study of heavy ion collisions and explosive astrophysical scenarios can 
complement each other in probing the phase diagram of strongly interacting matter, also in regard to the phase transition from hadronic to quark matter. 
However, the study of possible observable signals and impacts of quark matter in astrophysical systems requires 
hydrodynamical simulations with an input of an appropriate quark-hadron equation of state.
In this article we present such an approach where a quark matter phase 
transition has been implemented in a hadronic equation of state for a large range of temperatures, proton fractions and densities. 
Applying the latter to simulations of core-collapse supernovae, we find that a quark matter phase transition can cause the formation of a second shock wave 
which leads to the 
explosion of the star, accompanied by a second neutrino burst. The latter is
dominated by anti-neutrinos, which can be observed by future and present neutrino detectors. If found, the second neutrino peak can give correlated 
information about the progenitor mass and the critical density for the onset of quark matter. 
\paragraph{Acknowledgement}
The project was funded by the Swiss National Science Foundation grant. no. PP00P2-
124879/1 and 200020-122287 and the Helmholtz Research School for Quark Matter Studies,
the Italian National Institute for Nuclear Physics, the Graduate Program for Hadron and Ion
Research (PG-HIR), the Alliance Program of the Helmholtz Association (HA216/EMMI) and
the DFG through the Heidelberg Graduate School of Fundamental Physics. 
The work of G.~P. is supported by the Deutsche Forschungsgemeinschaft
(DFG) under Grant No. PA 1780/2-1. The authors are
additionally supported by CompStar, a research networking program of the European Science
Foundation, and the Scopes project funded by the Swiss National Science Foundation grant.
no. IB7320-110996/1. 

~

\end{document}